\documentclass[10pt, a4paper]{article}
\usepackage{amssymb,eurosym,amsfonts,amsmath,calc}
\usepackage{array}
\usepackage{indentfirst}
\usepackage{graphicx}
\usepackage{hyperref}
\usepackage{authblk}
\usepackage{color}
\usepackage{multicol}
\usepackage[latin9]{inputenc}
\usepackage{fancybox}
\usepackage[normalem]{ulem}

\begin{document}
\title{Resonant superalgebras for supergravity}
	\author{Remigiusz Durka \thanks{remigiusz.durka@uwr.edu.pl}}
	\author{Krzysztof M. Graczyk \thanks{krzysztof.graczyk@uwr.edu.pl}}
\affil{Institute for Theoretical Physics, University of Wroc\l{}aw, pl.\ M.\ Borna 9, 50-204 Wroc\l{}aw, Poland}
\setcounter{Maxaffil}{0}
\renewcommand\Affilfont{\itshape\small}

\maketitle


\begin{abstract}
Considering supergravity theory is a natural step in the development of gravity models. This paper follows the ``algebraic`` path and constructs possible extensions of the Poincar\'e and Anti-de-Sitter algebras, which inherit their basic commutation structure. Previously achieved results of this type are fragmentary and show only a limited fraction of possible algebraic realizations. 
Our paper presents the newly obtained symmetry algebras, evaluated within an efficient pattern-based computational method of generating the so-called 'resonating' algebraic structures. These supersymmetric extensions of algebras, going beyond the Poincar\'e and Anti-de Sitter ones, contain additional bosonic generators $Z_{ab}$ (Lorentz-like), and $U_a$ (translational-like) added to the standard Lorentz generator $J_{ab}$ and translation generator $P_{a}$. Our analysis includes all cases up to two fermionic supercharges, $Q_{\alpha}$ and $Y_{\alpha}$.
The delivered plethora of superalgebras includes few past results and offers a vastness of new examples. The list of the cases is complete and contains all superalgebras up to two of Lorentz-like, translation-like, and supercharge-like generators $(JP+Q)+(ZU+Y)=JPZU+QY$. In the latter class, among $667$ founded superalgebras, the $264$ are suitable for direct supergravity construction. For each of them, one can construct a unique supergravity model defined by the Lagrangian. As an example, we consider one of the algebra configurations and provide its Lagrangian realization.
\end{abstract}




\section{Introduction}

General Relativity (GR) has been the subject of various fundamental extensions and generalizations. One of the remarkable realizations embodies the idea of the supersymmetry between bosons and fermions. The supersymmetry joins a bosonic spin-$2$ (graviton) field with fermionic spin $3/2$ (Rarita-Schwinger) field to form the so-called supergravity (SUGRA) \cite{Freedman:1976xh, VanNieuwenhuizen:1981ae, Townsend:1977qa, Derendinger:2015cja}.

The corresponding supergravity action formulation, exploiting supersymmetry transformation interchanging bosons with fermions and vice versa, is built from the vielbein $e^{a}$, spin connection $\omega^{ab}$, and the Rarita-Schwinger spinor $\psi$:
\begin{align}\label{sugra-action}
I_{\text {sugra }} &=\frac{1}{64 \pi G}\int d^4x\, \left(R_{\mu \nu}^{a b} (\omega)\,e_{\rho}^{c} e_{\sigma}^{d}+\frac{1}{\ell^{2}} e_{\mu}^{a} e_{\nu}^{b} e_{\rho}^{c} e_{\sigma}^{d}\right) \epsilon_{a b c d} \epsilon^{\mu \nu \rho \sigma}\notag\\
&+\int d^4x\, \left(\frac{1}{2} \bar{\psi}_{\mu} \gamma_{5} \gamma_{a} e_{\nu}^{a} \mathcal{D}_{\rho}^{\omega} \psi_{\sigma}+\frac{1}{4 \ell} \bar{\psi}_{\mu} \gamma_{5} \gamma_{a b} e_{\nu}^{a} e_{\rho}^{b} \psi_{\sigma}\right) \epsilon^{\mu \nu \rho \sigma}\,.
\end{align}
It is based on the underlining structure of the superalgebra formed by the generators of local gauge symmetries: translations $P_a$, Lorentz transformations $J_{a b}$ and supercharge $Q_\alpha$. There are two established types of algebra closing, leading to two separate actions. They correspond to either the supersymmetric extensions of the AdS (Anti-de Sitter) (with $\Lambda=-\frac{3}{\ell^2}$) or the Poincar\'e (with $\ell\to \infty$) algebras. 

A further natural development was to enlarge the number of fermionic charges. However, in recent years, an effort has been made to include new generators on the bosonic part. The first such example, the so-called Maxwell algebra, originates from the past work of \cite{Schrader:1972zd, Bacry:1970ye}. D.~Soroka and V.~Soroka  have introduced another case \cite{Soroka:2011tc}, leading altogether to an interesting generalization called the \textit{semigroup expansion} \cite{Izaurieta:2006zz, Edelstein:2006se, Diaz:2012zza, Inostroza:2018gzd, Andrianopoli:2013ooa, Salgado:2014qqa, Hoseinzadeh:2014bla, Concha:2016kdz}. With a different approach, some similar results have been obtained by several other groups~\cite{Lukierski:2010dy, deAzcarraga:2010sw, Durka:2011nf}.

The semigroup expansion delivered a consistent way of proceeding with the task of finding other algebraic examples~\cite{Concha:2016hbt}, eventually leading to the notion of the \textit{resonant algebras}~\cite{Durka:2016eun, Durka:2019guk, Durka:2019vnb}. 

If decomposition of the semigroup satisfies the same structure as the sub-spaces of the original algebra, then the semigroup expansion is called \textit{resonant}. In this work, we fully exploit \textit{resonant} character of algebraic structures, understood as obeying the same structure constants pattern of the supersymmetric extension of the AdS (super)algebra, even for the extended content of the (super)generators. In the rest of the paper, we shall call such algebraic structures as the \textit{resonant algebras}.

Besides the bosonic discussion within the semigroup/resonant framework, the emphasis has also been placed on finding fermionic extensions. Recent examples correspond to Refs.~\cite{Soroka:2006aj, Bonanos:2010fw, Kamimura:2011mq, Durka:2011gm, Durka:2012wd, deAzcarraga:2014jpa,Concha:2018jxx} and \cite{Concha:2014tca, Concha:2015woa} but they are limited to just a few explicit cases for a particular content of generators.  

The program of computational brute-force finding of all bosonic algebras within the {semigroup/resonant} framework was concluded in Ref. \cite{Durka:2019vnb}. In the following paper \cite{Concha:2020atg}, using  a bit different approach based on \textit{generator} considerations, the supersymmetric $JPZ+Q$ configurations were obtained (with one additional Lorentz-like generator $Z_{a b}$). 

Additional generators in the bosonic and fermionic sectors make further analysis of the algebras extremely difficult because of an overwhelming number of variants to check. Indeed, we estimate that the computational framework (\textit{Cadabra} environment \cite{Peeters:2018dyg}) used in \cite{Concha:2020atg} needs about $27$ hours and two months of running time on one CPU to analyze $JPZ+QY$ and $JPZU+QY$ scenarios, respectively, where $U_{a}$ and $Y_\alpha$ correspond to a new translation-like and supercharge-like generators.

In this paper, we present a significantly improved method for getting resonant algebras, which for $JPZU+QY$ configuration, reduces the time of analysis from two months to one day. The pattern-based algorithm is implemented in \textit{Wolfram Mathematica} language. To fasten analysis, it exploits the stochastic procedures known in Monte Carlo simulations. The approach allows us to obtain the complete list of the possible resonant superalgebras, up to two fermionic generators, including only a few results known so far.

Indeed, thanks to the new algorithm, starting from $JP+Q$, we consider further configurations until the scheme is doubled, namely $(JP+Q)+(ZU+Y)=JPZU+QY$. Therefore, we present all cases up to two fermionic charges $Q_{\alpha}$ and $Y_{\alpha}$ in the presence of additional bosonic generators, $Z_{ab}$ or $U_a$, added to the standard Lorentz generator $J_{ab}$ and translation generator $P_{a}$ with all intermediate configurations. These superalgebras can be directly used for the construction of supergravity actions. 
In particular, the superalgebras $JPZU+QY$ give us a possibility to discuss bi-supergravity. Among $667$ founded supersymmetric extensions, for $264$ the relation $\{Q,Q\}\sim P+ \dots$ is satisfied.

Eventually, we discuss the impact of the resonant algebras structures on the action content. We also consider one of the $JPZU+QY$ superalgebras, using it to construct the three-dimensional (3D) supergravity.

In the following sections, we will give the basis of the framework resulting from the required conditions and explain details of the designed searching method, interesting by its own merits. 
We will show the algebraic outline and finish by providing some details of the supergravity construction. This paper comes with supplementary materials containing all the found algebras explicitly.


\section{From AdS algebra to resonant superalgebras}

We start with the Lie algebra of the Lorentz generators, $J$'s, which satisfy the commutation relations
\begin{align}
	\left[J_{ab},J_{cd}\right] & =\eta_{bc}J_{ad}-\eta_{ac}J_{bd}-\eta_{bd}J_{ac}+\eta_{ad}J_{bc}\, ,
	\label{equ1} 
\end{align}
where $a,b=0,1,2,...$ are the group indices, and $\eta_{ab}$ is the Minkowski metric. Adding translations requires
\begin{align}
	\left[ J_{ab}, P_{c}\right] & =\eta_{bc}P_{a}-\eta_{ac} P_{b}\,,  \label{equ2}.
\end{align}
To close above algebra three different variants are considered, namely, either: 
\begin{align}
	\left[ P_{a},P_{b}\right] =0\qquad \text{or}\qquad \left[ 
	P_{a},P_{b}\right] =J_{ab}\qquad \text{or} \qquad %
	\left[ P_{a},P_{b}\right]=-J_{ab}\,. \label{equ3}
\end{align}
These three scenarios correspond to the Poincar\'{e} $ISO(D-1,1)$, Anti de-Sitter (AdS) $SO(D-1,2)$, and de-Sitter (dS)  $SO(D,1)$ group of symmetries, respectively. The cosmological constant defined as $\Lambda=\mp \frac{3}{\ell^2}$ results from the re-scaling of generators in $\left[\ell P_{a}, \ell P_{c}\right]=\pm J_{ac}$. The conventions used throughout this paper, due to possible supergravity applications, will account only for the first two cases (Poincar\'{e} and AdS) upon whose other structures will be based.

Restricting to the negative cosmological constant (AdS), the above commutation relations can be written in the form of
\begin{align}
	\left[J_{a b}, J_{c d}\right]=f_{ab,cd}{}^{m s} J_{m s}, \qquad
	\left[J_{a b}, P_{c}\right]=f_{ab,c}{}^{m} P_{m}, \qquad
	\left[P_{a}, P_{b}\right]=f_{a,b}{}^{ms} J_{m s}\,,
\end{align}
where structure constants are defined as
\begin{align}
f_{ab,cd}{}^{m s} =-\delta_{a b}^{k m} \eta_{k n} \delta_{c d}^{n s},\qquad f_{ab,c}{}^{m}=-\delta_{a b}^{k m} \eta_{k n} \delta_{c}^{n}, \qquad f_{a,b}{}^{ms}=\delta_{ab}^{ms}\,.
\end{align}

The S-expansion \cite{Izaurieta:2006zz, Edelstein:2006se, Diaz:2012zza, Salgado:2014qqa} supply an interesting approach to systematically construct further bosonic algebras. That generalization having origins in the concept of the algebraic contraction \cite{Inonu:1953sp} represents the product of $S \times \mathfrak{g}$. The new generators are provided by the semigroup 'scaling':
\begin{align}
	J_{ab,(i)}=s_{2i}\tilde{J}_{ab}\qquad \text{and}\qquad P_{a,(i)}=s_{2i+1}\tilde{P}_{a},\qquad \text{for}~i=\{0,1,2,...\}\\
	\!\!\!\!\!\!\!\!\!\!\!\!\!\!\mathrm{where}\quad J_{ab}= s_0\tilde{J}_{ab},\qquad  P_{a}= s_1\tilde{P}_a,\qquad Z_{ab}= s_2 \tilde{J}_{ab},\qquad U_{a}= s_3\tilde{P}_a, \quad \dots 
\end{align}
with the original algebra $\mathfrak{g}=AdS$ given by $\tilde{J}_{ab},\tilde{P}_{a}$ and some semigroup elements $s_i \in S$ obeying the so-called resonant condition \cite{Izaurieta:2006zz, Diaz:2012zza, Salgado:2014qqa}. Mathematically, for the bosonic part, this could be seen a parity requirement:
\begin{equation}
	s_{even}\cdot s_{even}=s_{even}\,,\qquad s_{even}\cdot s_{odd}=s_{odd}\,,\qquad s_{odd}\cdot s_{odd}=s_{even},
\end{equation}
reflecting the AdS algebra structure 
\begin{equation}
[\tilde{J}_{..}, \tilde{J}_{..}]\sim \tilde{J}_{..}\qquad [\tilde{J}_{..}, \tilde{P}_{.}]\sim \tilde{P}_{.} \qquad [\tilde{P}_{.}, \tilde{P}_{.}]\sim \tilde{J}_{..}
\end{equation}

Differently to the standard  In\"{o}n\"{u}-Wigner contraction \cite{Inonu:1953sp}, we do not have some limit describing the transition from AdS to Poincar\'{e} as cosmological constant vanishes by $\ell \to \infty$. Instead, there is just an evaluation of the products determining the appearance of the new generators. At the same time, zero in the commutators is carried by the process of so-called $0_S$ reduction or simply including zero element as the semigroup absorbing element \cite{Durka:2016eun}.  

As the semigroup only changes the type of generator, the $S \times \mathfrak{g}$ product assures inheriting the structure constants, which are built with the same scheme as the starting $\mathfrak{g}=AdS$. We later exploit this feature directly also in the supersymmetric generalization. 

The semigroup expansion is called resonant when the decomposition of the semigroup satisfies the same structure as the sub-spaces of the original (super)algebra. In practice, to construct the so-called resonant (super)algebras, we start from (super)AdS with generators $JP(Q)$. Then we include the generators $J$-like, $P$-like, and $Q$-like type, keeping the same pattern of structure constants as for $JPQ$. Our framework can be then seen as simply filling out all possibilities depending on the generator content:
\begin{align}\label{empty}
	\left\{
	\begin{aligned} 
		\left[ \framebox(10,10){~~}_{ab},\ \framebox(10,10){~~}_{cd}\right] & =\eta_{bc} \framebox(10,10){~~}_{ad}-\eta_{ac}\ \framebox(10,10){~~}_{bd}-\eta_{bd} \framebox(10,10){~~}_{ac}+\eta_{ad} \framebox(10,10){~~}_{bc}\, ,\\
		\left[\framebox(10,10){~~}_{ab},\ \dashbox(10,10){~~}_{c}\right] & =\eta_{bc} \dashbox(10,10){~~}_{a}-\eta_{ac} \dashbox(10,10){~~}_{b}\, , \\
		\left[\dashbox(10,10){~~}_{a},\ \dashbox(10,10){~~}_{b}\right] & =\framebox(10,10){~~}_{ab}\, ,\\
		\left[ \framebox(10,10){~~}_{ab},\ovalbox{\phantom{I}}_{\alpha }\right]
		&=\frac{1}{2}\,\left( \Gamma _{ab}\right) _{\alpha }^{\beta }\ovalbox{\phantom{I}}_{\beta}\,,\\ 
		\left[ \dashbox(10,10){~~}_{a},\ \ovalbox{\phantom{I}}_{\alpha }\right]&=\frac{1}{2}\,\left( \Gamma _{a}\right) _{\alpha }^{\beta }\ovalbox{\phantom{I}}_{\beta}\,, \\ 
		\left\{ \ovalbox{\phantom{I}}_{~\alpha },\ \ovalbox{\phantom{I}}_{~\beta }\right\} &=-\left(	\Gamma ^{a} C\right) _{\alpha \beta } \dashbox(10,10){~~}_{~a}+\frac{1}{2}\left(\Gamma ^{ab} C\right) _{\alpha \beta } \framebox(10,10){~~}_{~ab}\,.
	\end{aligned} 
	\right.
\end{align}
Available content of generators concerns 
\begin{align}
	& \framebox(10,10){~~}_{~ab} \to J_{ab}, Z_{ab}, \ldots \quad\; (\rm{Lorentz-like})\notag\\
	& \dashbox(10,10){~~}_{~a}\to P_{a}, U_{a}, \ldots \quad\quad\, (\rm{translation-like})\\
	& \ovalbox{\phantom{I}}_{~\alpha } \to Q_\alpha, Y_\alpha, \dots \quad\quad (\rm{supercharge-like})\notag
\end{align}
and can be carried out further to include even more extended generator content. Above we introduce the Dirac matrices $\Gamma_a$, as well as $\Gamma_{ab}=1/2(\Gamma_a \Gamma_b-\Gamma_b \Gamma_a)$ and charge conjugation matrix $C$, whereas $a,b,c,d$ are Lorentz indices and $\alpha,\beta$ are spinorial indices. 

Naturally, not all possibilities are physically viable; therefore, we introduce a certain number of requirements. We also allow zero to appear as the output of the (anti)commutators on the right-hand sides of equations \eqref{empty}.

The algebraic and physical requirements which must be satisfied are the following: 
\begin{itemize}
	\item holding the same structure constants as original super AdS;
	\item preservation by the Lorentz generator, i.e. for all generators $[J,X]\sim X$;
	\item anticommutator $\{Q,Q\}$ being non-zero; 
	\item fulfilling graded super-Jacobi identities.
\end{itemize}
Above constraints assure algebraic structures suitable for constructing actions but mind that some would be considered exotic. For the proper supergravity construction, we demand additional condition:
\begin{equation}
\label{Eq_QQSugra}
\{Q,Q\}\sim P+\dots
\end{equation}  

Note that our framework, although containing an additional fermionic charge, does not include the central extension \cite{Gocanin:2019ota,Concha:2019icz,Caroca:2019dds,Concha:2020tqx}. Such extension allows us to encode $U(1)$ generator $T$ into the framework as the output of anticommutation of $Q$ with $Y$. We leave such considerations for future work. In this work, we allow two fermionic charges to appear similarly to Refs.~\cite{Lukierski:2010dy, Soroka:2006aj, Durka:2011gm, Durka:2012wd}. 
Note that in the three-dimensional case with $a,b=0,1,2$ we can introduce dual generators like $J_{a}=\frac{1}{2}\epsilon_{a}{}^{bc}\,J_{bc}$, that offer possibility of rewriting AdS uniformly as
\begin{align}
	\left[ J_{a},J_{b}\right] =\epsilon_{abc}J^{c},\qquad
	 \left[ J_{a},P_{b}\right] =\epsilon _{abc}P^{c},\qquad
	 \left[ P_{a},P_{b}\right] =\epsilon _{abc}J^{c}\,.
\end{align}
Including a fermionic supercharge $Q_\alpha$ gives us two superalgebra options: Poincar\'e and AdS respectively
\begin{align}
	\left\{ \begin{aligned} \left[ J_{a},J_{b}\right] &=\epsilon
		_{abc}J^{c}\,,\\ \left[ J_{a},P_{b}\right] &=\epsilon _{abc}P^{c}\,, 
		\\ \left[ P_{a},P_{b}\right] &=0\,, \\ \left[ J_{a},Q_{\alpha
		}\right] &=\frac{1}{2}\,\left( \Gamma _{a}\right) _{\alpha }^{\beta }Q_{\beta
		}\,, \\ \left[ P_{a},Q_{\alpha }\right]
		&=0\,, \\ \left\{ Q_{\alpha },Q_{\beta }\right\} &=-\left(
		\Gamma^{a}C\right) _{\alpha \beta }P_{a}\,  \end{aligned} \right.
	\qquad\qquad \left\{ \begin{aligned} \left[ J_{a},J_{b}\right] &=\epsilon
		_{abc}J^{c}\,,\\ \left[ J_{a},P_{b}\right] &=\epsilon _{abc}P^{c}\,, 
		\\ \left[ P_{a},P_{b}\right] &=\epsilon _{abc}J^{c}\,,  \\ \left[
		J_{a},Q_{\alpha }\right] &=\frac{1}{2}\,\left( \Gamma _{a}\right) _{\alpha
		}^{\beta }Q_{\beta }\,, \\ \left[ P_{a},Q_{\alpha }\right]
		&=\frac{1}{2}\,\left( \Gamma _{a}\right) _{\alpha }^{\beta }Q_{\beta
		}\,,\text{ \ \ }  \\ \left\{ Q_{\alpha },Q_{\beta }\right\}
		&=-\left(\left( \Gamma^{a} C\right) _{\alpha \beta } P_{a} + (\Gamma ^{a}C )_{\alpha \beta} J_{a} \right)\,.
	\end{aligned} \right.\label{3d}
\end{align}
It is an obvious realization of the super-Jacobi identities, where we leave out the trivial examples with the Lorentz generator:
\begin{align}
	[[P_a,P_b],J_c]+[[P_b,J_c],P_a]+[[J_c,P_a],P_b]&=0\,,\label{Jacobi3d-a} \\
	[[P_a,P_b],Q_\alpha]+[[P_b,Q_\alpha],P_a]+[[Q_\alpha,P_a],P_b]&=0\,, \label{Jacobi3d-b}\\
	\{[P_a,Q_\alpha],Q_\beta\}+[\{Q_\alpha,Q_\beta\},P_a]-\{[Q_\beta,P_a],Q_\alpha\}&=0\,, \label{Jacobi3d-c}\\
	[\{Q_\lambda,Q_\alpha\},Q_\beta]+[\{Q_\alpha,Q_\beta\},Q_\lambda]+[\{Q_\beta,Q_\lambda\},Q_\alpha]&=0\,.\label{Jacobi3d-d}
\end{align}
To better see the structure of Poincar\'{e} and AdS laid out in \eqref{3d}, let's rewrite it schematically as
\begin{align}\label{Poincare}
	\begin{tabular}[t]{c|cc}
		\lbrack .,.] & $J$ & $P$ \\ \hline
		$J$ & $J$ & $P$ \\ 
		$P$ & $P$ & $0$
	\end{tabular}
	\qquad 
	\begin{tabular}[t]{c|c}
		\lbrack .,.] & $Q$ \\ \hline
		$J$ & $Q$ \\ 
		$P$ & $0$
	\end{tabular}
	\qquad 	
	\begin{tabular}[t]{c|c}
		\{.,.\} & $Q$ \\ \hline
		$Q$ & $P$
	\end{tabular}
	\end{align}
and
\begin{align}
	\begin{tabular}[t]{c|cc}\label{AdS}
		\lbrack .,.] & $J$ & $P$ \\ \hline
		$J$ & $J$ & $P$ \\ 
		$P$ & $P$ & $J$
	\end{tabular}
	\qquad 
	\begin{tabular}[t]{c|c}
		\lbrack .,.] & $Q$ \\ \hline
		$J$ & $Q$ \\ 
		$P$ & $Q$
	\end{tabular}
	\qquad 
	\begin{tabular}[t]{c|c}
		\{.,.\} & $Q$ \\ \hline
		$Q$ & $P+J$
	\end{tabular}
\end{align}

That way of presenting algebras turns out to be very convenient. Not only it comes in a concise form, but it immediately highlights all the differences between algebras. It also emphasizes the independence of the form of structure constants. Depending on a given subject and particular applications, we can use either notation of AdS structure constants from \eqref{empty} in a general case or \eqref{3d} for $3D$. 


\section{Dynamical searching for resonant (super)algebras}

The main idea of our approach is the following, having the set of the given generators, \{$X_i \} = \{J_{ab}\,, P_{a},\, Q_\alpha \dots \}$, we postulate the algebra tables (such as in \eqref{Poincare} and \eqref{AdS}), with the action being super-commutator denoted by $(X_i,X_j)$. For the valid algebra all the super-Jacobi identities must be satisfied.
Note that the number of algebra candidates ((anti)commutation tables as an input) grows tremendously with the growing set of generators. At the same time, the number of super-Jacobi identities also grows as the number of generators increases. 

In the previous work \cite{Concha:2020atg}, all (anti)commutation relations were explicitly encoded and evaluated using $\Gamma$'s identities (like $\Gamma_{a b} \Gamma_{c}=\eta_{c b} \Gamma_{a}-\eta_{c a} \Gamma_{b}$) and the explicit representations, which was extremely time and resource consuming. Below, due to the \textit{resonant} character of the discussed (super)algebras, we point out significant simplification that omits the mathematical evaluation of particular structure constants allowing us to perform a highly efficient computer check disregarding wrong candidates.


\subsection{Super-Jacobi identity}

Certainly, for the correct algebra example the super-Jacobi identity is satisfied
\begin{align}
    \label{defJacobi}
    \mathrm{Jacobi}(X,Y,Z) =0\,, 
\end{align}
where 
\begin{align}
	\mathrm{Jacobi}(X,Y,Z) \equiv \left( \left(X_i, X_j\right), X_k\right)  \oplus \left(\left( X_j, X_k\right), X_i \right)  \oplus \left(\left( X_k, X_i\right), X_j\right)\,.
\end{align}
We use here $i,j$ as the any pair of indices attached to the various types of generators, and $\oplus$ sign for accounting $(-1)$ factor in the graded relation.

If one explicitly use the structure constants, then the super-Jacobi identity reads
\begin{align}
	0&=\color{red}\left( \color{black}\left(X_i, X_j\right), X_k\color{red}\right)\color{black}  \oplus \color{blue}\left( \color{black}\left( X_j, X_k\right), X_i\color{blue}\right)\color{black}  \oplus \color{green}\left(\color{black}\left( X_k, X_i\right), X_j\color{green}\right)\color{black} \notag\\
	&=f_{i,j}{}^{m}\color{red}\left( \color{black}X_m, X_k\color{red}\right)\color{black}  \oplus f_{j,k}{}^{m}\color{blue}\left( \color{black}X_m, X_i\color{blue}\right)\color{black}  \oplus f_{k,i}{}^{m}\color{green}\left(\color{black}X_m, X_j\color{green}\right)\color{black} \notag\\
	&=f_{i,j}{}^{m}\, f_{m,k}{}^{n}\color{red}\,\underbrace{X_n}_{A}\color{black} \oplus
	\color{black}f_{j,k}{}^{m}\, f_{m,i}{}^{n}\color{blue}\, \underbrace{X_n}_{B}   \color{black}\oplus f_{k,i}{}^{m}\,f_{m,j}{}^{n}\color{green} \underbrace{X_n}_{C}\color{black}\,.\label{sJI}
\end{align}

Based on the fundamental properties of starting $JP+Q$ algebra and resonant construction, each of three pieces $A$, $B$, $C$ of the expression \eqref{sJI} must be expressed by the same generator.  Moreover, the structure constants in \eqref{sJI} are determined by the original $JP+Q$ and curried further. Therefore, to satisfy the Jacobi identity $A=B=C$, including the case $A=B=C=0$. For $A \neq B$, we get linear dependence of $A$, $B$, $C$, and some generators being forced to be a linear combination of others, which is not the case for a considered list of uniquely defined generators. In the end, all that matters is a final consistency and matching of the final generator coming from the three pieces of the super-Jacobi identity by obtaining in the last step the same letter or get three zeros.

Before we go any further, let's point out other certain subtleties. After an evaluation of three fermionic generators in the super-Jacobi identity, we observe
\begin{align}
	0 & =(-(\Gamma^{a}C)_{\alpha\beta}(\Gamma_{a})_{\rho\lambda}-\frac{1}{4}%
	(\Gamma^{ab}C)_{\alpha\beta}(\Gamma_{ab})_{\rho\lambda})\ovalbox{\phantom{I}}^{~\rho} \notag\\
	& +(-(\Gamma^{a}C)_{\beta\lambda}(\Gamma_{a})_{\rho\alpha}-\frac{1}{4}%
	(\Gamma^{ab}C)_{\beta\lambda}(\Gamma_{ab})_{\rho\alpha})\ovalbox{\phantom{I}}^{~\rho} \notag\\
	& +(-(\Gamma^{a}C)_{\lambda\alpha}(\Gamma_{a})_{\rho\beta}-\frac{1}{4}%
	(\Gamma^{ab}C)_{\lambda\alpha}(\Gamma_{ab})_{\rho\beta})\ovalbox{\phantom{I}}^{~\rho} \,.
\end{align}
After gathering of terms
\begin{align}
	0=- &  \Big((\Gamma^{a}C)_{\alpha\beta}(\Gamma_{a})_{\rho\lambda}+(\Gamma
	^{a}C)_{\beta\lambda}(\Gamma_{a})_{\rho\alpha}+(\Gamma^{a}C)_{\lambda\alpha
	}(\Gamma_{a})_{\rho\beta}\Big)\ovalbox{\phantom{I}}^{~\rho} \notag\\
	&  -\frac{1}{4}\Big((\Gamma^{ab}C)_{\alpha\beta}(\Gamma_{ab})_{\rho\lambda
	}+(\Gamma^{ab}C)_{\beta\lambda}(\Gamma_{ab})_{\rho\alpha}+(\Gamma
	^{ab}C)_{\lambda\alpha}(\Gamma_{ab})_{\rho\beta}\Big)\ovalbox{\phantom{I}}^{~\rho}\,,
\end{align}
we can show that the first and second lines (each holding three contributions) separately vanish by substituting explicit gamma matrices.  
It also should be kept in mind while looking at (anti)commutator tables that the schematic outcome of $P+J$ corresponds to $\Gamma^a P_a+\Gamma^{ab} J_{ab}$, and so on. With these out the way, let us now focus on the implementation of the searching tool.


\subsection{Algorithm}

In this section, we formulate the algorithm based on the results of the previous sections. 
Having a given set of generators, we create all possible tables of commutators and anticommutators obeying the requirements mentioned in the previous sections. Each table defines (super)algebra candidate, denoted by $\mathrm{Alg}_i$. Then to determine whether it represents correct (super)algebra, one should check the fulfillment of the (super)Jacobi identities.  
But we have shown that (\ref{defJacobi}) is satisfied if 
\begin{equation}
	((X_i,X_j),X_k) \cong ((X_j,X_k),X_i) \cong ((X_k,X_i),X_j)\,,
\end{equation}
where $\cong$ means that two sides of equation are equal up to the sign. Note that $\mathrm{Jacobi}(X_i,X_j,X_k)\cong \mathrm{Jacobi}(X_i,X_k,X_j)$ and so on. Also obviously $(X_i, X_j) \cong (X_j,X_i)$, and $(X_i, 0) = 0$.

Having a given set of (super)algebra generators, we generate: 
\begin{itemize}
	\item set $Alg = \{Alg_1, \dots, Alg_{\overline{Alg}}\}$ containing all possible algebra candidate configurations, where $Alg_i$ is the list of (anti)commutation rules defining given candidate of algebra, and by $\overline{Alg}$ we denote the total number of candidates within the $Alg$ set
	\item all necessary Jacobi identities  $Jac=\{Jacobi_1, \dots  Jacobi_{\overline{Jacobi}}\}$,  where ${\overline{Jacobi}}$ denotes the number of Jacobi identities and $Jacobi_i = \mathrm{Jacobi}(X_k,X_m,X_n)$. 
\end{itemize}
Both numbers $\overline{Alg}$ and $\overline{Jacobi}$ depend on the particular generator content. 

The first property we must keep while generating valid candidates is
\begin{equation}
    (J, X_i) \sim X_i\,,
\end{equation}
i.e., Lorenz generator with any other, to preserve the Lorentz invariance. 

Additionally, the generators can be divided into three subsets: even indexed bosonic (like $J_{ab},Z_{ab},...$), odd indexed bosonic (like $P_a,U_a,...$), fermionic (like $Q_{\alpha}, Y_{\alpha},...$). Used even/odd separation has its direct roots in semigroup expansion with even/odd number labels of the semigroup elements, which transits into bosonic generators with two/one one group indices, respectively.

The resonant (super)algebras fulfills the following super-commutation relations:
\begin{eqnarray}
	\label{Eq:rule1}
	(\mathrm{even},\mathrm{even} ) & \sim & \mathrm{even}\quad \mathrm{or}\quad 0 \\ 
	\label{Eq:rule2}
	(\mathrm{even},\mathrm{odd} ) & \sim & \mathrm{odd}\quad \mathrm{or}\quad 0 \\
	\label{Eq:rule3}
	(\mathrm{odd},\mathrm{odd} ) & \sim & \mathrm{even}\quad \mathrm{or}\quad 0 \\
	\label{Eq:rule4}
	(\mathrm{boson},\mathrm{fermion} ) & \sim & \mathrm{fermion}\quad \mathrm{or}\quad 0 \\
	\label{Eq:rule5}
	(\mathrm{fermion},\mathrm{fermion} ) & \sim & \mathrm{even} + \mathrm{odd} \quad \mathrm{or}\quad \mathrm{even}, \quad \mathrm{or} \quad \mathrm{odd} \,.
\end{eqnarray}
Note, we allow  $\{Q,Y\}$ and $\{Y,Y\}$ to vanish, but we neglect the possibility of zero from $\{Q,Q\}$.

Now assuming that algebra contains: $p+1$, $n$ and $f$ of even indexed bosonic generators, odd indexed, and fermionic ones, respectively, then the generated total number of candidates configurations reads
\begin{equation}
	\overline{Alg} = \underbrace{(p+2)^{ \frac{p(p+1)}{2}}}_{\mathrm{from}\,(\ref{Eq:rule1})} \underbrace{(n+1)^{p n}}_{\mathrm{from}\,(\ref{Eq:rule2})} \underbrace{(p+2)^{\frac{n(n+1)}{2}}}_{\mathrm{from}\,(\ref{Eq:rule3})} \underbrace{(f+1)^{(p+n)f}}_{\mathrm{from}\,(\ref{Eq:rule4})} \underbrace{((p+2)(n+1)-1)((p+2)(n+1))^{\frac{(f+1)f}{2}-1}}_{\mathrm{from}\,(\ref{Eq:rule5})}.
\end{equation}

The next step is to generate all Jacobi identities which must be fulfilled by successful algebra candidate. To form single Jacobi identity we choose three generators from the given set of generators. However, $\mathrm{Jacobi}(X_i,X_i,X_i)$ doesn't offer any new information. Also if  $J$ is one of these generators then Jacobi identity is automatically fulfilled. Indeed,
\begin{equation}
	\mathrm{Jacobi}(X_i,X_j,J) =   ((X_i,X_j),J) \oplus ((X_j,J),X_i) \oplus ((J,X_i),X_j) \cong  (X_i,X_j) \oplus (X_i,X_j) \oplus (X_i,X_j)\,.
\end{equation}
Thanks to above, the number of generators used to obtain necessary Jacobi identities equals $p+n+f$. From this set we must choose three elements though combinations with repetition, so the total number of unique identities to check reads: 
\begin{equation}
	\overline{\mathrm{Jacobi}} = \begin{pmatrix} p+n+f+2 \cr 3\end{pmatrix}.
\end{equation}

Then having given candidate $Alg_i$ with a unique set of (anti)commutation rules, the upper number of checks to execute is equal $6\cdot \overline{\mathrm{Alg}}\,\cdot \,\overline{\mathrm{Jacobi}}$ . Factor six accounts for two rounds of super-commutator evaluations in a single super-Jacobi identity as shown in \eqref{sJI}. For \textit{Cadabra}, in the case of $JPZU+QY$, it would be unavoidable to perform $35$ unique super-Jacobi identities multiplied by six super-commutator substitutions for each of $344\,373\,768$ possible algebra candidates. Therefore, we naturally turn towards \textit{Mathematica} and non-standard searching described below.

The algorithm consists of several steps, namely:
\begin{enumerate}
		\item Consider given set of generators. Then following rules, explained above, form the set of candidate algebras, $Alg$, 
		as well as a set of Jacobi identities, $Jac$.  
		\item  Let $i=1$, $k=1$, where ($i=1, \dots$~$\overline{\mathrm{Alg}}$), ($k=1,\dots$, $\overline{\mathrm{Jacobi}}$). 
		\item \label{step00}
		Consider algebra $\mathrm{Alg}_i$   and then
		\begin{enumerate} 
			\item \label{step_10} consider the $k$-th Jacobi identity, $Jacobi_k$, and check its correctness. If test is: 
			\begin{enumerate}
				\item \label{step_20} positive and  $k = \overline{\mathrm{Jacobi}}$ then  the algebra is saved and go to the step (\ref{step_11}) else $ k \to k+1$ and go to step (\ref{step_10});
				\item \label{step_21} negative, then stop verification, save the Jacobi identity for which the test failed and go to step (\ref{step_11}).
			\end{enumerate}
			\item \label{step_11} If $i$~mod~$10$ is zero then shuffle the order of Jacobi identities and go to next step. 
			
			\item \label{step_12} Set $i \to i+1$, $k\to 1$
			and if $i<\overline{\mathrm{Alg}}$ go to the step (\ref{step00}) else stop. 
		\end{enumerate}
\end{enumerate}
\begin{figure}[h!]
\centering
        \includegraphics[width=0.85\textwidth]{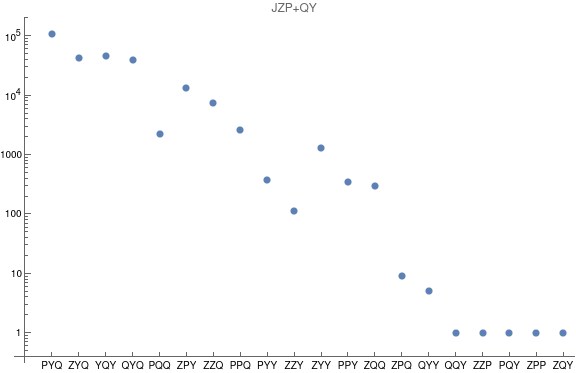}
\caption{The final distributions of the most 'problematic' Jacobi identities during the search process. "PQQ" refers to $\mathrm{Jacobi}(P,Q,Q)$ and so on.}
    \label{fig:disttributions}
\end{figure}

Notice that during the process of verification,
by registering the problematic Jacobi identities, we construct the distribution, histogram, of the most troublesome Jacobi's identities for fulfillment. 
Then in every $10$-steps of $i$-loop, we re-order the Jacobi identities in $Jac$ by drawing the order from the reconstructed distribution. 
It makes the algorithm partially stochastic and significantly accelerates the process of searching for the correct algebraic structures. 
This way, we avoid many trivial identities being satisfied by most of the algebra candidates.

Summarizing: the algorithm ''learns'' the most problematic Jacobi identities and uses them in the search process. 
In Fig.~\ref{fig:disttributions} we show the final histogram obtained for $JPZ+QY$ algebra. We see that for the algorithm, the most critical identities to verify are $\mathrm{Jacobi}(P,Y,Q)$, $\mathrm{Jacobi}(Z,Y,Q)$, $\mathrm{Jacobi}(Y,Q,Y)$, and $\mathrm{Jacobi}(Q,Y,Q)$, etc. 


\section{Overview of superalgebras}

Eventually, closing of the supersymmetric algebras can be achieved in many distinctive ways, according to the provided table:
\begin{table}[h!]
	\label{table-superalgebras}
	\begin{center}
		\begin{tabular}{l|l|l}
			~1$\times$ J             & ~1$\times$ J+Q             & ~~0$\times$ J+QY             \\ \hline
			~2$\times$ JP            & ~2$\times$ JP+Q            & ~\textbf{10}$\times$ JP+QY    \\ \hline
			~6$\times$ JPZ    & ~9$\times$ JPZ+Q  & \textbf{102}$\times$ JPZ+QY  \\ \hline
			30$\times$ JPZU & \textbf{43}$\times$ JPZU+Q &\textbf{667}$\times$ JPZU+QY
		\end{tabular}
		\caption[]{Resonant algebras and superalgebras depending on the generator content.}
	\end{center}
\end{table}

Starting $J$ configuration is the Lorentz algebra, and two mentioned examples of $JP$ configurations obviously correspond to the Poincar\'e and AdS algebras. Including supercharge leads to $JP+Q$ Poincar\'e and AdS superalgebras. If we continue extending the generator content, we eventually reproduce known examples and find a plethora of new (super)algebras. We emphasize in bold the complete listings of the newly obtained configurations reported in this paper.

To assure the construction of the supergravity models, we demand $\{Q, Q\}= P+...$ (with or without further contribution). This constraint reduces significantly obtained cases. Such examples we will call resonant supergravity algebras:
\begin{table}[h!]
	\label{table-supergravityalgebras}
	\begin{center}
		\begin{tabular}{l|l}
			~2$\times$ JP+Q & ~~~8$\times$ JP+QY    \\ \hline
			~6$\times$ JPZ+Q  & ~~71$\times$ JPZ+QY  \\ \hline
			17$\times$ JPZU+Q & 264$\times$ JPZU+QY
		\end{tabular}
		\caption[]{Possible resonant superalgebras to construct supergravity.}
	\end{center}
\end{table}

\subsubsection*{Resonant J, JP, JPZ, JPZU algebras}

As mentioned above, $J$ is the Lorentz algebra, whereas $JP$ contains Poincar\'e and AdS. The $JPZ$ and $JPZU$ cases are bosonic resonant algebras one can find in \cite{Durka:2016eun, Durka:2019guk, Durka:2019vnb}. Among them we obtain well known examples like Maxwell algebra $\mathfrak{B}_{4}$ \cite{Schrader:1972zd, Bacry:1970ye} and $\mathfrak{C}_{4}$ given by Soroka-Soroka~\cite{Soroka:2006aj,Durka:2011nf,Salgado:2014qqa}, as well as their generalization to $\mathfrak{B}_{m}$ and $\mathfrak{C}_{m}$ families.

\subsubsection*{Resonant J+Q, JP+Q, JPZ+Q superalgebras}

Configuration $J+Q$ is simply super Lorentz. The next example, $JP+Q$, offers $\mathcal{N}=1$ supergravity formulation in the form of the supersymmetric extensions of the Poincar\'e and AdS algebras. Configurations with the additional Lorentz-like generator $Z_{ab}$, namely $JPZ+Q$, appeared recently \cite{Concha:2020atg}, although here we point out the corrected number of examples \footnote{Present reevaluation result of \cite{Concha:2020atg} has showed a typo in one of the tables carried from the original article of \cite{Durka:2016eun}. That has an impact on the correct outcome of the branch claimed not to have a supersymmetric counterpart. Instead, we find that there is one more consistent supersymmetric extension with one spinor charge for the $B_{4}$ to be included in prepared errata of \cite{Concha:2020atg}
\begin{equation*}
	\begin{array}{llllllll}
		B_{4}: &  &
		\begin{tabular}[t]{c|ccc}
			\lbrack .,.] & $J$ & $P$ & Z \\ \hline
			$J$ & $J$ & $P$ & Z \\
			$P$ & $P$ & 0 & $P$ \\
			Z & Z & $P$ & $J$
		\end{tabular}
		&  &
		\begin{tabular}[t]{c|c}
			\lbrack .,.] & $Q$ \\ \hline
			$J$ & $Q$ \\
			$P$ & 0 \\
			Z & $Q$
		\end{tabular}
		&  &
		\begin{tabular}[t]{c|c}
			\{.,.\} & $Q$ \\ \hline
			$Q$ & $P$
		\end{tabular}
		& ~~~~~~
	\end{array}
\end{equation*}
} being 9. That accounts to 6 SUGRA examples and 3 non-standard (exotic) cases for which $\{Q,Q\}$ gives something else than $P$. Adopting names from \cite{Concha:2020atg}, we have exactly one SUGRA case for Soroka-Soroka $\mathfrak{C}_4$, as well as $B_4$, and $C_4$, whereas  $\tilde{B}_4$ has 2 SUGRA versions + 1 non-standard, $\tilde{C}_4$ has 1 SUGRA + 1  non-standard, and the Maxwell case $\mathfrak{B}_4$ delivers only 1  non-standard example.

\subsubsection*{Resonant JPZU+Q superalgebras}

For the first time, we derive here the class of $JPZU+Q$, where we deal with the supersymmetric extension for the generator content equipped with additional Lorentz-like $Z_{ab}$ and translational-like $U_{a}$ generators. Thirty bosonic resonant algebras $JPZU$ have 43 supersymmetric extensions with 17 suitable cases for the valid supergravity formulation. For the details, we send the reader to the attached supplemented file.

\subsubsection*{Resonant JP+QY superalgebras}

In this paragraph, we present findings containing two fermionic supercharge generators $Q$ and $Y$. Configurations $JP+QY$ come with ten total examples, among them 8 SUGRA and two non-standard exotic cases. Below we present them organized as three AdS-like and seven Poincar\'e-like ones:
$$
\begin{array}[c]{l|ll}
  & J & P \\
 \hline
 J & J & P \\
 P & P & J \\
\end{array}\qquad\qquad\begin{array}[c]{l|ll}
  & Q & Y \\
 \hline
 J & Q & Y \\
 P & Y & Q \\
\end{array}\qquad
\left\{\quad
\begin{array}[c]{l}
 \begin{array}[t]{l|ll}
  & Q & Y \\
 \hline
 Q & P+J & P+J \\
 Y & P+J & P+J \\
\end{array}
\\~~\\
\begin{array}[t]{c|cc}
  & Q & Y \\
 \hline
 Q & P & J \\
 Y & J & P \\
\end{array}
\\~~\\
\begin{array}[t]{c|cc}
  & Q & Y \\
 \hline
 Q & J & P \\
 Y & P & J \\
\end{array}
\end{array}
\right.
$$ 
$$
\begin{array}[c]{lll}
\begin{array}[c]{l|ll}
  & J & P \\
 \hline
 J & J & P \\
 P & P & 0 
\end{array}
\quad
\begin{array}{r}
\left\{
\begin{array}{l}
\\[195pt]
\end{array}
\right.\\
\\[10pt]
\end{array}
\begin{array}[c]{l}
\begin{array}[c]{c|ll}
  & Q & Y \\
 \hline
 J & Q & Y \\
 P & Y & 0 \\
\end{array}\qquad
\left\{\quad
\begin{array}[c]{l}
 \begin{array}[t]{l|cl}
  & Q & Y \\
 \hline
 Q & P+J & P \\
 Y & P & 0 \\
\end{array}
\\~~\\
\begin{array}[t]{c|cc}
  & Q & Y \\
 \hline
 Q & P & 0 \\
 Y & 0 & 0 \\
\end{array}
\\~~\\
\begin{array}[t]{c|cc}
  & Q & Y \\
 \hline
 Q & J & P \\
 Y & P & 0 \\
\end{array}
\end{array}
\right.\\~~\\
\begin{array}[c]{c|cc}
  & Q & Y \\
 \hline
 J & Q & Y \\
 P & 0 & 0 \\
\end{array}\qquad
\left\{\quad
\begin{array}[c]{l}
 \begin{array}[t]{c|cc}
  & Q & Y \\
 \hline
 Q & P & P \\
 Y & P & P\\
\end{array}
\\~~\\
\begin{array}[t]{c|cc}
  & Q & Y \\
 \hline
 Q & P & P \\
 Y & P & 0 \\
\end{array}
\\~~\\
\begin{array}[t]{c|cc}
  & Q & Y \\
 \hline
 Q & P & 0 \\
 Y & 0 & P \\
\end{array}
\\~~\\
\begin{array}[t]{c|cc}
  & Q & Y \\
 \hline
 Q & P & 0 \\
 Y & 0 & 0 \\
\end{array}
\end{array}
\right.
\end{array}
\end{array}\qquad
$$

Interestingly, there is  In\"{o}n\"{u}-Wigner contraction scheme with the parameter $\sigma$ scaling the generators:
\begin{align}
J_{ab}\to \sigma^0 J_{ab}, \quad P_{a}\to \sigma^1 P_{a}, \quad Z_{ab}\to \sigma^2 Z_{ab}, \quad U_{a}\to \sigma^3 U_{a},\quad
Q_{\alpha}\to \sigma^{\frac{1}{2}} Q_{\alpha}, \quad Y_{\alpha}\to \sigma^{\frac{3}{2}} Y_{\alpha}\,.
\end{align}
In the $\sigma\to\infty$ limit, we can directly relate three AdS-like superalgebras with the first three Poincar\'{e}-like, leaving a branch of last four Poincar\'{e}-like cases separated.

\subsubsection*{Resonant JPZ+QY superalgebras}

For the $JPZ+QY$ resonant superalgebras, there are 102 examples based on the six bosonic tables $JPZ$ with the 'family' names once again adopted from \cite{Concha:2020atg}. Extensive search shows that we have:	
\begin{itemize}
	\item $\mathfrak{C}_4$: 8 total (with 7 leading to SUGRA + 1  non-standard) [Soroka-Soroka]
	\item $\mathfrak{B}_4$: 7 total (with 2 leading to SUGRA + 5  non-standard) [Maxwell]
	\item $\tilde{C}_4$: 14 total (with 9 leading to SUGRA + 5  non-standard)
	\item $\tilde{B}_4$: 57 total (with 38 leading to SUGRA + 19  non-standard)
	\item $C_4$: 10 total (with 9 leading to SUGRA + 1  non-standard)
	\item $B_4$: 6 total (with 6 leading to SUGRA + 0  non-standard)\,.
\end{itemize}

\subsubsection*{Resonant JPZU+QY superalgebras}

The last analyzed scenario concerns 667 examples of closed superalgebras with effectively two sets of Lorentz-like, translational-like, supercharge-like generators. These supersymmetric extensions coming from thirty $JPZU$ bosonic resonant algebraic families, a priori require over $10^{10}$ checks (achieved within one day instead of estimated two months of direct \textit{Cadabra} symbolic calculations).

Only 264 are suitable to construct the supergravity as they have $\{Q, Q\} = P+...$. Among them, we would like to point out only three that restore supersymmetric AdS sub-algebra with $[P,P]\sim J$ and $\{Q, Q\} \sim P+J$, which might turn out interesting in some applications. They are
\begin{align}
\begin{array}[t]{c|cccc}
		\framebox{$\mathcal{B}_{5}$} & J & P & Z & U \\ \hline
		J & J & P & Z & U\\
		P & P & J & U & Z\\
		Z & Z & U & 0 & 0\\
		U & U & Z & 0 & 0
\end{array}
\qquad
\begin{array}[t]{c|cc}
	& Q & Y \\ \hline
	J & Q & Y \\
	P & Q & Y \\
	Z & Y & 0 \\
	U & Y & 0 \\
\end{array}
\qquad
\begin{array}[t]{c|cc}
	& Q & Y \\ \hline
	Q & P+J & U+Z \\
	Y & U+Z & 0 \\
\end{array}
\\~~\nonumber\\
\begin{array}[t]{c|cccc}
		\framebox{$\mathcal{D}_{5}$} & J & P & Z & U \\ \hline
		J & J & P & Z & U\\
		P & P & J & U & Z\\
		Z & Z & U & Z & U\\
		U & U & Z & U & Z
	\end{array}
\qquad
\begin{array}[t]{c|cc}
	& Q & Y \\ \hline
	J & Q & Y \\
	P & Q & Y \\
	Z & Y & Y \\
	U & Y & Y \\
\end{array}
\qquad
\begin{array}[t]{c|cc}
	& Q & Y \\ \hline
	Q & P+J & U+Z \\
	Y & U+Z & U+Z \\
\end{array}
\\~~\nonumber\\
\begin{array}[t]{c|cccc}
		\framebox{$\mathcal{C}_{5}$}  & J & P & Z & U \\ \hline
		J & J & P & Z & U\\
		P & P & J & U & Z\\
		Z & Z & U & J & P\\
		U & U & Z & P & J
	\end{array}
\qquad
\begin{array}[t]{c|cc}
	& Q & Y \\ \hline
	J & Q & Y \\
	P & Q & Y \\
	Z & Y & Q \\
	U & Y & Q \\
\end{array}
\qquad
\begin{array}[t]{c|cc}\label{mathcalC5}
	& Q & Y \\ \hline
	Q & P+J & U+Z \\
	Y & U+Z & P+J \\
\end{array}
\end{align}
For convenience we write $\mathcal{C}_5$ superalgebra \eqref{mathcalC5} in its explicit form:
\begin{align}
		\left[J_{ab},J_{cd}\right] & =\left[Z_{ab},Z_{cd}\right] =\eta_{bc} J_{ad}-\eta_{ac} J_{bd}-\eta_{bd} J_{ac}+\eta_{ad} J_{bc}\, ,\notag\\
		\left[J_{ab},Z_{cd}\right] & =\eta_{bc} Z_{ad}-\eta_{ac} Z_{bd}-\eta_{bd} Z_{ac}+\eta_{ad} Z_{bc}\, ,\notag\\
		\left[J_{ab},P_{c}\right] & =\left[Z_{ab},U_{c}\right]=\eta_{bc} P_{a}-\eta_{ac} P_{b}\, , \notag\\
		\left[J_{ab},U_{c}\right] & =\left[Z_{ab},P_{c}\right]=\eta_{bc} U_{a}-\eta_{ac} U_{b}\, , \notag\\
		\left[P_{a},P_{b}\right] & =\left[U_{a},U_{b}\right] =J_{ab}\, ,\notag\\
		\left[P_{a},U_{b}\right] & = Z_{ab}\, ,\notag
		\end{align}
\begin{align}		
		\left[ J_{ab},Q_{\alpha }\right]
		&=\left[ Z_{ab},Y_{\alpha }\right]=\frac{1}{2}\,\left( \Gamma _{ab}\right) _{\alpha }^{\beta }Q_{\beta}\,,\notag\\
		\left[ J_{ab},Y_{\alpha }\right]
		&=\left[ Z_{ab},Q_{\alpha }\right]=\frac{1}{2}\,\left( \Gamma _{ab}\right) _{\alpha }^{\beta }Y_{\beta}\,,\notag\\
		\left[P_{a},Q_{\alpha}\right]&=\left[U_{a},Y_{\alpha}\right]=\frac{1}{2}\,\left( \Gamma_{a}\right)_{\alpha }^{\beta }Q_{\beta}\,,\notag\\
		\left[P_{a},Y_{\alpha}\right]&=\left[U_{a},Q_{\alpha}\right]=\frac{1}{2}\,\left( \Gamma_{a}\right)_{\alpha }^{\beta }Y_{\beta}\,, \notag\\
		\left\{ Q_{\alpha },Q_{\beta}\right\} &=\left\{ Y_{\alpha },Y_{\beta}\right\} =-\left(	\Gamma ^{a} C\right) _{\alpha \beta } P_{a}+\frac{1}{2}\left(\Gamma ^{ab} C\right)_{\alpha \beta } J_{ab}\,, \notag\\ 
		\left\{ Q_{\alpha },Y_{\beta}\right\}  &=-\left(	\Gamma ^{a} C\right) _{\alpha \beta } U_{a}+\frac{1}{2}\left(\Gamma ^{ab} C\right) _{\alpha \beta } Z_{ab}\,.\label{mathcalC5explicit}
\end{align}

The rest of the $JPZU+QY$ tables can be found in the supplemented file.


\section{Resonant supergravities and bi-supergravity}

For each of the superalgebras shown in the previous section, one can construct a distinct supergravity model defined by the appropriate Lagrangian. The necessary elements of this construction are the following: (i) the gauge connection one-form~$\mathbb{A}$; (ii) the super-curvature two-form~$\mathbb{F}$; (iii) the gauge parameter~$\Theta$, along with the gauge transformations; (iv) the Killing form $\left\langle \dots \right\rangle$.  As it goes beyond the scope of the present paper, we shall not provide a complete list of supergravity Lagrangians. We leave it for future work. However, we give some highlights concerning action construction base on the resonant superalgebras.

\subsection{Sub-invariant sectors from resonant superalgebras}

The Killing metric (form), used to contract all the group indices, takes the form of the invariant tensor given for any combination of two generators, i.e. $\left\langle J_{ab}\,J_{cd}\right\rangle =\alpha_{0}\,\epsilon_{abcd}$, $\left\langle P_{a}\,P_{b}\right\rangle =\alpha_{0}\,\eta_{ab}$, $\left\langle Q_\alpha\, Q_\beta\right\rangle=(\alpha_1-\alpha_0)\,C_{\alpha\beta}$, $\left\langle J_{ab}\,Z_{cd}\right\rangle =\alpha_{2}\epsilon_{abcd}$, and so on. Note that resonant/semigroup framework remarkably introduces sub-invariant sectors through the arbitrary valued $\alpha$'s constants in front of invariant tensors. The non-vanishing components for a given (super)algebra can be established from the known super-AdS outcomes and an identity $\langle[X_i , X_j ]\, X_k \rangle = \langle X_i \, [X_j , X_k ] \rangle$ with $X_i$ being any generator. They also can be taken from the semigroup expansion framework. Their appearance is directly associated with the form of the particular algebras \cite{Durka:2016eun, Durka:2019guk, Durka:2019vnb}, which leads to different term content and decomposition of terms between various sub-sectors. As an example lets consider $JPZ+Q$ configurations (with the transition to the dual 3D fields $\omega^a=\frac{1}{2}\epsilon^{abc}\omega_{bc}$, $h^a=\frac{1}{2}\epsilon^{abc}h_{bc}$ and generators definitions, $J_a=\frac{1}{2}\epsilon_{a}{}^{bc}J_{ab}$, $Z_a=\frac{1}{2}\epsilon_{a}^{bc}{}Z_{ab}$. Also we are going to introduce $\mathcal{R}^{a}=d\omega^{a}+\frac{1}{2}\epsilon^{abc}\omega_{b}\omega_{c}$, $\mathcal{D}_{\omega}e^{a}=d e^{a}+\epsilon^{abc}\omega_{b}\,e_{c}$,  the Lorentz covariant derivative $\mathcal{D}_{\omega }\psi=d\psi+\frac{1}{2}\omega^{a}\gamma_{a}\psi$, as well as define $\mathcal{F}=\mathcal{D}_{\omega }\psi+\frac{1}{2\ell}e^{a}\gamma_{a}\psi+\frac{1}{2}h^{a}\gamma_{a}\psi$. The most general three-dimensional $\mathcal{N}=1$ CS supergravity action 
\begin{equation}
I_{CS}=\frac{k}{4\pi }\int_{\mathcal{M}}\left\langle \mathbb{A} \wedge d\mathbb{A}+\frac{1}{3} \mathbb{A}\wedge [\mathbb{A},\mathbb{A}]\right\rangle \,,  \label{CSaction}
\end{equation}
 being invariant under the Soroka-Soroka $\mathfrak{C}_{4}$ superalgebra reads
\begin{eqnarray}
	I_{CS}^{\mathfrak{C}_{4}} &=&\frac{k}{4\pi }\int \left[ \alpha _{0}\left(
	\omega ^{a}d\omega _{a}+\frac{1}{3}\epsilon^{abc}\omega _{a}\omega
	_{b}\omega _{c}\right) \right.  \notag \\
	&&\left. +\alpha _{1}\left( \frac{2}{\ell}\mathcal{R}^{a}e_{a}+\frac{1}{3\ell^3}\epsilon
	^{abc}e_{a}e_{b}e_{c}+\frac{2}{\ell}e_{a}D_{\omega }h^{a}+\frac{1}{\ell}\epsilon ^{abc}e_{a}h_{b}h_{c}+\frac{2}{\ell}
	\bar{\psi}\mathcal{F} \right) \right.  \label{CSmathfrak_C4}  \\
	&&\left. +\alpha _{2}\left(\frac{1}{\ell^2}e_{a}D_{\omega }e^{a}+2h_{a}\mathcal{R}
	^{a}+\frac{1}{\ell^2}\epsilon^{abc}e_{a}e_{b}h_{c}+h_{a}D_{\omega }h^{a}+\frac{1}{3}\epsilon ^{abc}h_{a}h_{b}h_{c}-\frac{2}{\ell}\bar{\psi}\mathcal{F} \right) \right] \,.  \notag
\end{eqnarray}
At the same time, for the Maxwell algebra $\mathfrak{B}_4$ CS action is:
\begin{eqnarray}\label{CSmathfrak_B4} 
	I_{CS}^{\mathfrak{B}_{4}} &=&\frac{k}{4\pi }\int \left[ \alpha _{0}\left(
	\omega ^{a}d\omega _{a}+\frac{1}{3}\epsilon ^{abc}\omega _{a}\omega
	_{b}\omega _{c}\right) +\alpha _{2}\left(\frac{1}{\ell^2}
	e_{a}D_{\omega }e^{a}+2h_{a}\mathcal{R}^{a}-\frac{2}{\ell}\bar{\psi}\mathcal{D}_{\omega }\psi\right)\right.  \notag \\
	&&\left. +\alpha _{1}\left( \frac{2}{\ell}\mathcal{R}^{a}e_{a}\right)  \right] \,,
\end{eqnarray}
whereas the $B_4$ case, mention in earlier section, leads to the action of the form:
\begin{eqnarray}
	I_{CS}^{B_{4}} &=&\frac{k}{4\pi }\int \left[ \alpha _{0}\left(
	\omega ^{a}d\omega _{a}+\frac{1}{3}\epsilon ^{abc}\omega _{a}\omega
	_{b}\omega _{c}+h_{a}D_{\omega}h^{a}\right)  +\alpha _{2}\left(2h_{a}\mathcal{R}
	^{a}+\frac{1}{3}\epsilon ^{abc}h_{a}h_{b}h_{c} \right) \right.  \notag \\
	&&\left. +\alpha _{1}\left( \frac{2}{\ell}\mathcal{R}^{a}e_{a}+\frac{2}{\ell}e_{a}D_{\omega }h^{a}+\frac{1}{\ell}\epsilon ^{abc}e_{a}h_{b}h_{c}+\frac{2}{\ell}
	\bar{\psi}\mathcal{D}_\omega \psi+\frac{1}{\ell} \bar{\psi}h_{a}\gamma^{a} \psi \right) \right]\,.  \label{CS_B4} 
\end{eqnarray}

\subsection{Bi-supergravity}

We have obtained all superalgebras up to doubling the $JP+Q$ configuration. In the latter scenario the gauge one-form connection is being gauged over the $JPZU+QY$ generators:
\begin{equation}
\label{Eq:Aconnection}
	\mathbb{A}=\frac{1}{2} \omega^{ab}J_{ab}+\frac{1}{\ell} e^{a}P_{a}+\frac{1}{2} h^{ab}Z_{ab}+\frac{1}{\tilde{\ell}} k^{a} U_{a}+\frac{1}{\sqrt{\ell}}\psi^{\alpha}Q_{\alpha}+\frac{1}{\sqrt{\tilde{\ell}}}\chi^{\alpha}Y_{\alpha}\,,
\end{equation}
where $\omega^{ab}$ is the spin-connection one-form, $e^{a}$ is the vielbein, $h^{ab}$ is the additional gauge field related to $Z_{ab}$, $k^{a}$ related to $U_a$ and $\psi$ is the gravitino and $\chi$ yet additional gravitino. Note the presence of the two constants $\ell$ and $\tilde{\ell}$ to assure correct dimensions of fields. 

The super-curvature two-form is built straightforwardly from $\mathbb{A}$ connection, as $\mathbb{F}=d\mathbb{A}+\frac{1}{2}\left[\mathbb{A},\mathbb{A}\right]$:
\begin{equation}
	\mathbb{F}=\frac{1}{2} F^{ab}J_{ab}+\frac{1}{\ell} T^{a}P_{a}+\frac{1}{2} H^{ab}Z_{ab}+\frac{1}{\tilde{\ell}} K^{a} U_{a}+\frac{1}{\sqrt{\ell}}\mathcal{F}^{\alpha}Q_{\alpha}+\frac{1}{\sqrt{\tilde{\ell}}}\mathcal{G}^{\alpha}Y_{\alpha} \label{curvature-form-3D}
\end{equation}
with the particular parts depended on the chosen (super)algebra. 

The gauge parameter $\Theta	=\frac{1}{2} \lambda^{ab}J_{ab}+ \tau^{a}P_{a}+\frac{1}{2} \tilde{\lambda}^{ab}Z_{ab}+\tilde{\tau}^{a} U_{a}+\epsilon^{\alpha}Q_{\alpha}+\varepsilon^{\alpha}Y_{\alpha}$
allows us to write transformation law for the particular fields under the Lorentz $\lambda^{ab}$, translations $\tau^a$, so-called 'Maxwellian' transformations: $\tilde{\lambda}^{ab}$, $\tilde{\tau}^a$ and both supercharges $\epsilon^\alpha$, $\varepsilon^\alpha$. The transformation law $\delta_\Theta \mathbb{A}= \mathbb{D}^\mathbb{A}\Theta= d\Theta+[\mathbb{A},\Theta]$
describes particular laws uniquely determined by the explicit (super)algebra, with particular $\delta_\Theta\omega^{ab},\delta_\Theta e^{a},\delta_\Theta h^{ab},\delta_\Theta k^{a},\delta_\Theta\psi,\delta_\Theta\chi$ transformations. Remember also that $\delta_\Theta \mathbb{F}= [\Theta,\mathbb{F}]$. Depending on the construction, we can achieve the full gauge invariance (by $3D$ Chern-Simons theory \cite{Concha:2014tca,Concha:2015woa,Hassaine:2016amq,Concha:2021jos}) or in $4D$ just settle for the Lorentz and supersymmetry invariance with a possibility of additional invariance also due to the 'Maxwellian symmetries' \cite{Durka:2011gm,Durka:2012wd,Durka:2011va}.

To complete this section, we choose the superalgebra configuration $\mathcal{C}_5$ given in \eqref{mathcalC5}, which is interesting due to preserving sub-AdS superalgebra both in the commutator between two translations and in anticommutator of supercharges $\{Q,Q\} =  P+J$ . After using 3D dual definitions of fields ($\omega^a$ and $h^a$) and generators ($J_a$ and $Z_a$) we rewrite the connection $\mathbb{A}=\omega^{a}J_{a}+\frac{1}{\ell} e^{a}P_{a}+ h^{a}Z_{a}+\frac{1}{\tilde{\ell}} k^{a} U_{a}+\frac{1}{\sqrt{\ell}}\psi^{\alpha}Q_{\alpha}+\frac{1}{\sqrt{\tilde{\ell}}}\chi^{\alpha}Y_{\alpha}$ to obtain $\mathbb{F}=F^{a}J_{a}+\frac{1}{\ell} T^{a}P_{a}+ H^{a}Z_{a}+\frac{1}{\tilde{\ell}} K^{a} U_{a}+\frac{1}{\sqrt{\ell}}\mathcal{F}^{\alpha}Q_{\alpha}+\frac{1}{\sqrt{\tilde{\ell}}}\mathcal{G}^{\alpha}Y_{\alpha} $with the following components:
\small
\begin{align}
	F^{a}& =\mathcal{R}^{a}(\omega)+\frac{1}{2\ell^2}\epsilon^{abc} e_b e_c+\frac{1}{2}\epsilon^{abc}h_b h_c+\frac{1}{2\tilde{\ell}^2}\epsilon^{abc}k_b k_c+\frac{1}{2\ell}\,\bar{\psi}\Gamma^{a}\psi+\frac{1}{2\tilde{\ell}}\bar{\chi}\Gamma^{a}\chi\,,  \notag \\
	T^{a}& =D_\omega e^{a}+\frac{\ell}{\tilde{\ell}}\,\epsilon^{abc} h_{b} k_{c}+\frac{1}{2}\bar{\psi}\Gamma^{a}\psi+\frac{\ell}{2\tilde{\ell}}\bar{\chi}\Gamma ^{a}\chi  \,,  \notag \\
	H^{a}& =D_{\omega}h^{a}+\frac{1}{\ell\tilde{\ell}}\epsilon^{abc}e_b k_c+\frac{1}{2\sqrt{\ell\tilde{\ell}}}\,\bar{\psi}\Gamma^{ab}\chi+\frac{1}{2\sqrt{\ell\tilde{\ell}}}\,\bar{\chi}\Gamma^{ab}\psi\,,  \notag \\
	K^{a}& =  D_{\omega} k^{a}+\frac{\tilde{\ell}}{\ell}\epsilon^{abc}h_b e_c+\frac{1}{2}\sqrt{\tilde{\ell}/ \ell}\,\bar{\psi}\Gamma^{a}\chi+\frac{1}{2}\sqrt{\tilde{\ell}/ \ell}\,\bar{\chi}\Gamma^{a}\psi)\,,  \notag \\
\mathcal{F} & =\mathcal{D}_{\omega }\psi +\frac{1}{2\ell}\,e^{a}\Gamma_{a}\psi  +\frac{1}{2} \sqrt{\ell / \tilde{\ell}}\,h^{a}\Gamma_{a}\chi+\frac{1}{2}\sqrt{\ell / \tilde{\ell}^3}\,k^{a}\Gamma_{a}\chi\,,  \notag\\
\mathcal{G} & =\mathcal{D}_{\omega }\chi +\frac{1}{2\ell}e^{a}\Gamma _{a}\chi +\frac{1}{2} \sqrt{\tilde{\ell}/ \ell}\,h^{a}\Gamma _{a}\psi+\frac{1}{2}\frac{1}{\sqrt{\ell\tilde{\ell}}}\,k^{a}\Gamma_{a}\psi \,.
\end{align}\normalsize
Corresponding 3D Chern-Simons model $I_{CS}= \frac{k}{4\pi }\int \mathcal{L}$ for the algebra $\mathcal{C}_5$ has Lagrangian:
\small
\begin{eqnarray}
& &	\mathcal{L}^{\mathcal{C}_{5}}=\alpha _{0}\left(
	\omega ^{a}d\omega _{a}+\frac{1}{3}\epsilon^{abc}\omega _{a}\omega
	_{b}\omega _{c}+\frac{1}{\ell^2}e_{a}D_{\omega }e^{a}+h_{a}D_{\omega }h^{a}+\frac{1}{\tilde{\ell}^2}k_{a}D_\omega k^{a}+\frac{2}{\ell\tilde{\ell}}\epsilon^{abc}h_a e_b k_c-\frac{2}{\ell}
	\bar{\psi}\mathcal{F}-\frac{2}{\tilde{\ell}}
	\bar{\chi}\mathcal{G}\right)  \notag \\
	&&\left. +\alpha _{1}\left( \frac{2}{\ell}e_{a}\mathcal{R}^{a}+\frac{2}{\tilde{\ell}}k_a D_\omega h^a+\frac{1}{3\ell^3}\epsilon
	^{abc}e_{a}e_{b}e_{c}+\frac{1}{\ell}\epsilon^{abc}e_{a}h_{b}h_{c}+\frac{1}{\ell\tilde{\ell}^2}\epsilon^{abc}e_a k_b k_c+\frac{2}{\ell}
	\bar{\psi}\mathcal{F}+\frac{2}{\tilde{\ell}}
	\bar{\chi}\mathcal{G} \right) \right. \notag  \\
	&&\left. +\alpha _{2}\left(2h_{a}\mathcal{R}
	^{a}+\frac{2}{\ell \tilde{\ell}}e_a D_\omega k^a+\frac{1}{\ell^2}\epsilon^{abc}e_{a}e_{b}h_{c}+\frac{1}{3}\epsilon ^{abc}h_{a}h_{b}h_{c}+\frac{1}{\tilde{\ell}^2}\epsilon^{abc}h_a k_b k_c-\frac{2}{\sqrt{\ell\tilde{\ell}}}\bar{\psi}\mathcal{G}-\frac{2}{\sqrt{\ell\tilde{\ell}}}\bar{\chi}\mathcal{F} \right)  \right. \notag  \\
	&&+\alpha _{3}\left(\frac{2}{\tilde{\ell}}k_{a}\mathcal{R}
	^{a}+\frac{2}{\ell}e_{a}D_{\omega }h^{a}+\frac{1}{3\tilde{\ell}^3}\epsilon ^{abc}k_{a}k_{b}k_{c}+\frac{1}{\tilde{\ell}}\epsilon^{abc}h_{a}h_{b}k_{c}+\frac{1}{\ell^2\tilde{\ell}}\epsilon^{abc}e_{a}e_{b}k_{c}+\frac{2}{\sqrt{\ell\tilde{\ell}}}\bar{\psi}\mathcal{G}+\frac{2}{\sqrt{\ell\tilde{\ell}}}\bar{\chi}\mathcal{F} \right)  \,.
\end{eqnarray}\normalsize

The described above framework opens the possibility of bi-supergravity for $JPZU+QY$. Standardly, we assign the gravity formulation with the gauge theory with the spin connection $\omega^{ab}$ and vielbein $e^{a}$ altogether encoding the graviton field. Doubling of the field content in the form of additional $h^{ab}, k^{a}$ suitably offers grounds for the bi-metric theory \cite{Blas:2005yk, Banados:2008fi, Hoseinzadeh:2017lkh}. The connection $\mathbb{A}$ gauged for $JPZU+QY$ superalgebras enables the possibility of the bi-supergravity formulation with two sets of Lorentz-like, translation-like, and supercharge-like generators. Instead of the arbitrary mixing in the bi-metric action, see Ref.~\cite{Durka:2011va}, we can try to accompany the base metric (assigned to $J_{ab},P_{a}$) with other rank-2 field (corresponding to pair $Z_{ab},U_{a}$) and just follow particular superalgebra realization within the Chern-Simons construction in the $3D$ or $5D$ \cite{Concha:2016kdz}, whereas for $4D$ use the Born-Infeld type of action \cite{Concha:2014vka}, MacDowell-Mansouri model \cite{MacDowell:1977jt}, or the deformed BF theory \cite{Durka:2011nf, Durka:2012wd}. We leave evaluating the field equations and the ansatz solution discussion for future work.


\section{Summary}

Searching for the realization of supergravity theory should be based on the algebraic structure's proper choice describing the underlying symmetry. In this work, we fully exploit the 'resonant' character of algebraic structures, understood as obeying the same structure constants pattern of the supersymmetric extension of the AdS algebra. We proposed a new method of searching for the resonant (super)algebras. The new approach allowed us significantly go beyond the framework of Ref.~\cite{Concha:2020atg}. The advancement comes from the non-standard dealing with the Jacobi identities and developing a new (stochastic-like) searching algorithm. Thanks to our unique approach, we deliver the complete overview of all possible configurations starting from $JP+Q$ until the generator scheme is doubled, i.e., $JPZU+QY$. As a result, we provide a plethora of superalgebra structures that significantly complete the past fragmentary results. 

The detailed analysis of the obtained results definitely gives a broader perspective over the differences between various realizations of closing algebras and related features. For instance, we note the unique and not interchangeable role of the Lorentz generator that forces some of the constraints. The same can be said about the outcome of $\{Q, Q\}$. Indeed, it might be useful to bring forward some further commutation limitations on the possible configurations to restrict the superalgebra realizations, like non-vanishing of $\{Y, Y\}$. Eventually, we emphasize the direct applicability of our framework in action construction, including bi-supergravity. 


\section*{Acknowledgments}
The authors would like to thank prof. Jerzy Lukierski for valuable comments on the manuscript. The research project for both authors was partly supported by the program ''Excellence initiative - research university'' for the years 2020-2026 for the University of  Wroc\l{}aw.


\end{document}